\documentclass[11pt,preprint2]{aastex}
\usepackage{latexsym}
\usepackage{graphics}
\usepackage{enumerate}
\def\hide#1{}
\def\hh{h^{-1}}

\begin{document}

\def\dim#1{\mbox{\,#1}}
\def\HI{{\rm HI}}
\def\HII{{\rm HII}}
\def\lya{{Lyman-$\alpha$}}
\def\ll{{Lyman-limit}}

\title{Lyman Limit Systems in Cosmological Simulations}
\author{Katharina Kohler \altaffilmark{1,2}, Nickolay Y.\ Gnedin \altaffilmark{2,3}}

\altaffiltext{1}{JILA, University of Colorado, Boulder, CO 80309, USA} 
\altaffiltext{2}{Particle Astrophysics Center,
Fermi National Accelerator Laboratory, Batavia, IL 60510, USA; kkohler, gnedin@fnal.gov}
\altaffiltext{3}{Department of Astronomy \& Astrophysics, The
  University of Chicago, Chicago, IL 60637 USA}
\begin{abstract}
We used cosmological simulation with self-consistent radiative
transfer to investigate the physical nature of Lyman limit systems at
$z=4$. In agreement with previous studies, we find that most of \ll\
systems are ionized by the cosmological background, while higher
column density systems seem to be illuminated by the local sources of
radiation. In addition, we find that most of Lyman limit systems in
our simulations are located within the virial radii of galaxies with a
wide range of masses, and are physically associated with them (``bits
and pieces'' of galaxy formation). While the finite resolution of our
simulations cannot exclude an existence of a second population of
self-shielded, neutral gas clouds located in low mass dark matter
halos (``minihalos''), our simulations are {\it not\/} consistent with
``minihalos'' dominating the total abundance of Lyman limit systems.

\end{abstract}


\section{Introduction}

The hydrogen absorption system observed in spectra of distant quasars are
traditionally subdivided into \lya\ forest
($N_{\HI}\la1.6\times10^{17}\dim{cm}^2$), \ll\ systems
($1.6\times10^{17}\dim{cm}^2\la N_{\HI}\la10^{20}\dim{cm}^2$), and
damped \lya\ systems ($N_{\HI}\ga10^{20}\dim{cm}^2$). Amazingly, among
these three classes, the \ll\ systems are least understood, both
observationally and theoretically. This is quite unfortunate, given
that the \ll\ systems dominate the absorption of ionizing photons in
the universe \citep{me03}, and are, therefore, crucial for
understanding the transfer of ionizing radiation in the IGM and the
interactions between the early galaxies and quasars and their
environments.

Previous attempts to model \ll\ systems in cosmological simulations
identified them with low mass dark matter halos, filled with neutral,
self-shielded gas \citep{katz96,gard97,gard01}. On the other hand, the
observations of neutral hydrogen in the vicinity of the Milky Way
galaxy uncovered a large population of High Velocity Clouds (HVCs)
that could plausible be low redshift counterparts of \ll\ systems
\citep{o66,v69,blitz99,mal03,put03}. 

In this paper we use cosmological simulations with the self-consistent
treatment of the transfer of ionizing radiation to address the
question of the nature of \ll\ systems, as their appear in the simulations.
Another advantage of the simulations we use here is that they are
designed to model the reionization of the universe, and so achieve a
reasonable agreement with the SDSS observational data on the \lya\
forest at $z>5$ \citep{fsbw06,gf06}.

\section{Simulations}
\label{sec-simulations}

Simulations used in this paper have been run with the ``Softened
Lagrangian Hydrodynamics'' (SLH) code \citep{g00a,g04}. Simulations
include dark matter, gas, star formation, chemistry, and ionization
balance in the primordial plasma, and three-dimensional radiative
transfer.  The simulations described in this paper use a flat
$\Lambda$CDM cosmology with values of cosmological parameters as
determined by the first year WMAP data
\citep{svp03}. \footnote{$\Omega_M=0.27$, $\Omega_\Lambda=0.73$,
$h=0.71$, $\Omega_B=0.04$, $n_S=1$.}
                                                                               
We use two sets of simulations in this paper. The two sets differ by
the size of the computational box: in the first set the computational
box has the size of $4h^{-1}$ comoving Mpc, while in the second set
the box is $8h^{-1}$ comoving Mpc on a side.
                                                                               
The primary simulations in each of two sets are runs L4N128 and L8N128
from \citet{gf06}. Both of them include $128^3$ dark matter particles
and the same number of quasi-Lagrangian mesh cells for the gas
evolution, as well as a smaller number of stellar particles that
formed continuously during the simulation.
                                                                               
As \citet{gf06} emphasized, these simulations provide an adequate fit
to the evolution of the mean transmitted flux in the \lya\ forest
between $5<z<6.2$, but overproduce the opacity in the forest at
$z<5$. As we show below, they also overproduce \ll\ systems.
                                                                               
The simulations, however, contain adjustable parameters; the one of
importance here is the ionizing efficiency parameter $\epsilon_{UV}$,
that controls the amount of ionizing radiation emitted per unit mass
of stars formed in the simulation. This parameter depends on the
numerical resolution of the simulation and on the details of stellar
feedback and radiative transfer implementations, and has no clear
physical interpretation. It can be thought of as a fraction of
ionizing radiation escaping from the spatial scales unresolved in the
simulation (it has no direct relation to the widely used ``escape
fraction'' quantity).
                                                                               
Since we treat the ionizing efficiency as a free parameter, we can
adjust it to obtain a better agreement with the observational
data. For that purpose, we run additional simulations in which we
increased the ionizing efficiency by an additional factor $q$. Because
in this paper we only concentrate on comparing the simulations with
the data at $z=4$, and in order to save computational resources, we
only rerun the simulations with different values of the ionizing
intensity from $z=4.3$ to $z=4$. The redshift interval of $\Delta
z=0.3$ corresponds to the time interval of $130\dim{Myr}$, which is
some 4{,}000 larger than the mean photoionization time at this
redshift \citep{macdon02}, and, thus, should be more than
sufficient for establishing a new photoionization equilibrium even in
the lowest density regions. To be on the safe side, we also rerun one
of the simulations from $z=4.6$ to $z=4$, and the results for the
\lya\ mean transmitted flux and the column density distribution of the
\ll\ systems from that simulation are indistinguishable from the
analogous run started at $z=4.3$.
                                                                               
In order to make references to specific simulations transparent, we
label each simulation with a letter L followed by the value of the
linear size of the computational volume (measured in
$h^{-1}\dim{Mpc}$), followed by the letter ``q'' and the value of the
ionization enhancement parameter $q$. For example, the primary
simulations in each set are thus labeled L4q1 and L8q1, while a
$4h^{-1}\dim{Mpc}$ simulation with 6 times higher ionizing efficiency
is labeled L4q6. In the $4h^{-1}\dim{Mpc}$ set we run simulations
L4q1, L4q2, L4q6, and L4q10, while in the $8h^{-1}\dim{Mpc}$ set we
run L8q1 and L8q6.

The Lyman Limit systems are found by casting $10000$ lines of sight
through the simulation box, allowing us to sample the simulation box
finely enough to find even rare absorbers. The lines of sight are cast
in random directions and cover 
a redshift range of $\delta z = 0.033$ for the $8\hh\dim{Mpc}$
boxes. This narrow extent in redshift space makes it difficult to search
for the \ll\ edge in the spectra. Thus, to find the regions of high
optical depth along the lines of sight, we search for peaks in
$\lambda d\tau_{LL}/{d\lambda}$. 

This is similar to searching for an absorption line. For example, a
\lya\ absorption line profile can be described with (we ignore natural
width of the line here): 
\begin{equation}
\tau_{\alpha}=\frac{\tau_{0}}{b\sqrt{\pi}}\int
d\lambda'e^{-\frac{(\lambda-\lambda')^{2}}{b^{2}}}\delta(\lambda'-\lambda).
\end{equation}
When considering $\lambda d\tau_{LL}/d\lambda$ instead of $\tau_{LL}$,
the Lyman limit edge becomes the delta-function in frequency, in full
analogy with the absorption line profile. Thus, a \ll\ system appears
as a peak in the quantity
\begin{equation}
\tilde{\tau}_{LL}=\frac{1}{b\sqrt{\pi}}\int d\lambda'
e^{-\frac{(\lambda-\lambda')^{2}}{b^{2}}} \lambda'
\frac{d\tau_{LL}}{d\lambda'}.
\end{equation}

This substitution allows us to use software tools, already developed for
finding absorption lines in synthetic spectra, for finding \ll\ systems
in the simulated synthetic spectra. In particular, the
value of $\tilde{\tau}_{LL}$ at the ``line'' center is directly
proportional to the neutral hydrogen column density,
$\tilde{\tau}_{LL}=\sigma_{LL}N_{HI}$, where $\sigma_{LL}=6.3\times
10^{-18}\rm{cm^{2}}$ is the hydrogen photoionization cross-section. All
regions in the spectra where $\tilde{\tau}_{LL} \ge 1.0$,
corresponding to column 
densities of $N_{HI} > 1.6 \times 10^{17} \rm{cm^{-2}}$, are selected
as \ll\ systems.

After casting the lines of sight, the positions of these regions (in
3D space) and
their opacities are determined. We also compute other
characteristics such as local photoionization rate, column density and
neutral fraction of the gas. This information allows us to investigate
whether the systems are ionized by local sources or by the cosmological
background radiation.

Finding the positions of the \ll\ systems in the simulation box and
determining the position of the galaxies in the simulation allows us
to determine whether each Lyman Limit system can be associated with a
galaxy.

Galaxies in the simulation are identified with gravitationally bound
objects, found with the DENMAX algorithm \citep{bg91}. The DENMAX
algorithm works by constructing the finite resolution smooth total
(dark matter + gas + stars) density field from the simulation data. It
then identifies density maxima and associates them with bound
objects. This approach has a major advantage for our purposes here,
since it allows us to assign a specific meaning to the word
``associated''. Since each galaxy is represented by DENMAX as a
density hill, we will call all points in space that lie on that hill
as ``associated'' with that galaxy. In other words, a point of space
is ``associated'' with a given galaxy if, by moving against the
density gradient from that point one would eventually end up at the
center of the galaxy.
                                                                               
DENMAX separates all points of space into four distinct categories:
\begin{itemize}
\item points gravitationally bound to a given galaxy;
\item points associated with a given galaxy but not gravitationally bound to
it
(i.e.\ points lying on the galaxy density hill, but outside the virial
radius);
\item unassociated points (i.e.\ points not associated with any
  galaxy, like points in the middle of the void); and
\item points associated with unresolved galaxies (i.e.\ points
  belonging to the density hill too small to be qualified as a
  resolved galaxy).
\end{itemize}
We use this classification below to describe locations of \ll\ systems
in space.

\section{Results}
\label{sec-results}
The main goal of this work is to investigate the physical properties
of the Lyman limit systems themselves, such as neutral fraction and
photoionization rate, but also what kind of galaxies they are
associated with.

To begin, we have to normalize the amount of photons in the simulation
box to the observations, so that the population of simulated \ll\ systems
corresponds to the systems that are observed. This is done
using the number density of Lyman limit systems per column density and
adjusting the ionization efficiency to reduce the amount of \ll\
systems produced. Another observational parameter that can be used
to fit the simulations to observations is the mean transmitted flux in
Lyman-$\alpha$.

\begin{figure}[th]
\begin{center}
\epsscale{1.0}
\plotone{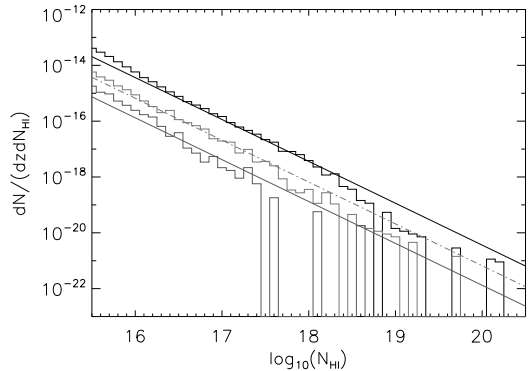}
\end{center}
\caption{Shown are the number densities of Lyman-Limit systems for
  three different runs, each with $4\hh\dim{Mpc}$ box size. The fits
  are overlaid. Top: L4q1, middle: L4q6, bottom: L4q10.}
\label{fig:Nsyshist4}
\end{figure}

Figure~\ref{fig:Nsyshist4} shows the number density of the Lyman limit
systems versus their column density for 3 different runs with varying
ionization efficiency $q$. The top run, with its fitted line overlaid,
corresponds to L4q1, which is the the run with the default value of
the ionization efficiency. The middle run is L4q6, where the
ionization efficiency is increased by a factor of 6 and the bottom run
is run L4q10, with a ionization efficiency $q=10$. The overlaid fits
are best-fit power laws with the slope $-1.5$ \citep{pet93,
me03}.

For the L4q6 run the fitted line exactly corresponds to the observed
number density obtained by \citet{S-L94}. The other two fits have the
same power law, but different amplitudes. L4q1 had a number density of
\ll\ systems too high to fit the observations. This illustrates that
the original simulations, while fitting the mean transmitted flux in
the \lya\ forest at $z>5$, significantly overpredict the amount of
\ll\ systems at $z=4$.

\begin{figure}[th]
\begin{center}
\epsscale{1.0}
\plotone{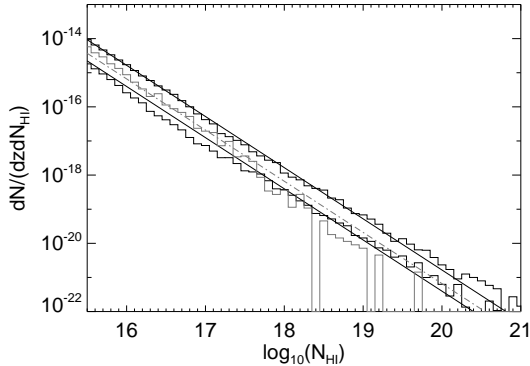}
\end{center}
\caption{Similarly to Figure~\ref{fig:Nsyshist4}, this graph shows
  the number density distribution of Lyman limit systems for two
  $8\hh\dim{Mpc}$ runs (in \textit{black}, L8q1 on top and L8q6 on
  bottom) and $4\hh\dim{Mpc}$ run L4q6 (in \textit{grey}). Their power
  law fits are included.}
\label{fig:Nsyshist8}
\end{figure}

Figure~\ref{fig:Nsyshist8} similarly shows the number density
distribution with column density for the two $8\hh\dim{Mpc}$ runs L8q1
and L8q6 and for comparison the $4\hh\dim{Mpc}$ run L4q6. The highest
distribution is again L8q1 with the lowest ionizing efficiency. The
other $8\hh\dim{Mpc}$ run also has the same $q=6$. There is good
agreement between the $4$ and the $8\hh\dim{MpC}$ run at
$N_{HI}>10^{17}\dim{cm}^2$. At lower column densities the larger box size lacks
enough resolution to fully resolve the \lya\ forest. Thus the number of
\lya\ forest systems is too low at lower column densities, which can be
seen from the disagreement for the two $q=6$ distributions.

\begin{figure}[th]
\begin{center}
\epsscale{1.0}
\plotone{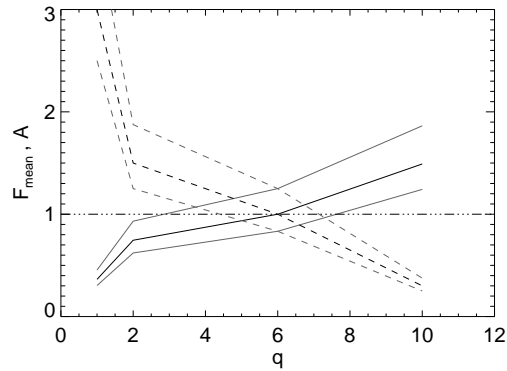}
\end{center}
\caption{This graph shows the ionization efficiency $q$ that was used
  as a fitting parameter for the simulations on the x-axis and shown on the
  y-axis are the two observational constraints: amplitude of the
  number density fit, A, and the mean transmitted flux $F_{mean}$. }
\label{fig:variables}
\end{figure}

In Figure~\ref{fig:variables} the dependence on the parameter $q$ is
illustrated. The solid lines show the mean transmitted flux in the
\lya\ forest normalized by the observational values of $0.55$ at
$z=4$. For higher $q$ there is more flux transmitted as the increased
ionizing intensity ionizes more of the initially neutral gas. The
dashed lines show the amplitude of the power law that is needed to fit
the number density of \ll\ systems as described above in
Figures~\ref{fig:Nsyshist4} and ~\ref{fig:Nsyshist8}. Here, for lower
ionization efficiency the offset is lower, for $q=6$ the offset is
unity, since the number density distribution is best fit by the
observational value. The horizontal line shows the observational
values for both parameters since we normalize by the observed
value. The gray lines correspond to a $20\%$ error, as is expected
from the observations. This plot illustrates that we can fit two
different observables with just one parameter, the ionization
efficiency, and that increasing this parameter by a factor of $6$ from
the values used in \citet{gf06} fits the observations at $z=4$.

\begin{figure}[th]
\begin{center}
\epsscale{1.0}
\plotone{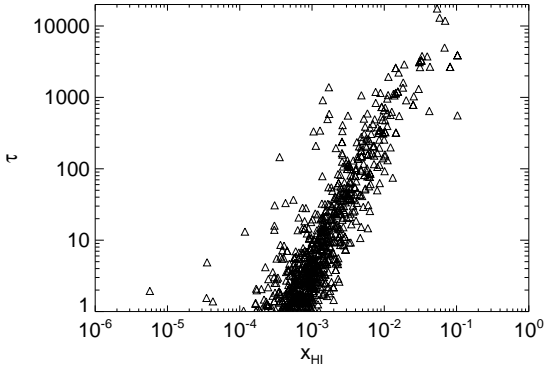}
\end{center}
\caption{Distribution of Lyman limit system optical depth versus the
  neutral fraction of gas.}
\label{fig:nfractau}
\end{figure}

Figure~\ref{fig:nfractau} shows the distribution of optical depth of
the \ll\ systems versus hydrogen neutral fraction. There is a
tight relation of the optical depth with neutral fraction, the higher
$\tilde{\tau}_{LL}$, the less ionized the system. The remarkable
conclusion from 
this is that the systems are substantially ionized and are not
neutral. This implies that the \ll\ 
systems detected in this simulation are not small self-shielded clumps
of neutral gas, but rather diffuse highly non-spherical clouds of gas
that happen to be oriented in their longest direction toward an
observer. 

\begin{figure}[th]
\begin{center}
\epsscale{1.0}
\plotone{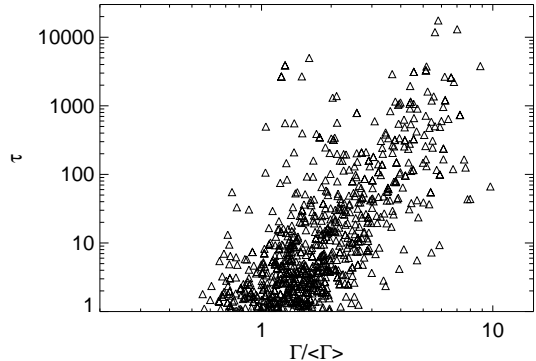}
\end{center}
\caption{Similar to Figure~\ref{fig:nfractau}, this graph shows the
  distribution of optical depth versus the normalized photoionization
  rate.}
\label{fig:gammatau}
\end{figure}

In Figure~\ref{fig:gammatau} it is similarly shown how the optical
depth depends on the photoionization rate for each \ll\ system. These
two variables are not as tightly correlated as the optical depth and
the neutral fraction. For lower $\tilde{\tau}_{LL}$ systems the normalized
photoionization rate $\Gamma/\langle\bar{\Gamma}\rangle$, where
$\bar{\Gamma}$ is the average $\Gamma$, is about unity. This means that these
systems are ionized by the cosmological background and not by local
sources. Systems with higher $\tilde{\tau}_{LL}$ appear to be ionized by local
sources because their normalized photoionization rate is higher. This
result is in agreement with previous findings by \citet{me05}and \citet{s04}.

One of the questions we investigate is whether the \ll\ systems are
associated with high or low mass galaxies. At first, we have tried to
use a method similar to one used by observational studies of the
environment of the \ll\ systems at low redshifts \citep{pent02}. In
observations, the \ll\ systems are often associated with a 
galaxy by finding the nearest object, either along a line of sight or
near it. A similar method was also used to analyze the simulations in
\citet{katz96}. There, a relation between the distance to the nearest
galaxy and $N_{\rm{HI}}$ of the \ll\ systems was shown, and it was concluded
that the column density 
correlates inversely with projected distance, and that \ll\ systems
either lie in the outer parts of massive protogalaxies or closer to
the center of less massive galaxies.

In our case we found that associating the \ll\ systems with the
nearest object is not necessarily the best option, because such
association is strongly dependent on the galaxy sample. When we decrease
the mass limit of the sample, thus including less and less massive
galaxies in the sample, we find progressively lower mass galaxies
closer and closer to a typical \ll\ system. This, of course, does not
imply that the \ll\ systems are physically associated with these low
mass galaxies, because low mass galaxies are more numerous, and the
distances between even a random subset of points in space to galaxies
of progressively smaller masses will be progressively smaller.

Thus, associating the \ll\ systems with nearest galaxies could lead to
the conclusion that most of the \ll\ systems are located in small
galaxies. However, using the physical association as permitted by the
DENMAX algorithm (\S \ref{sec-simulations}) - i.e.\ associating a \ll\
system with the galaxy whose ``density hill'' (i.e. dark matter halo)
the \ll\ is located on - we find a different result: the \ll\ systems
are associated with galaxies in a wide range of masses and the small
galaxies near them are often physically unrelated satellites of the
larger galaxy.

\begin{figure}[th]
\begin{center}
\epsscale{1.0}
\plotone{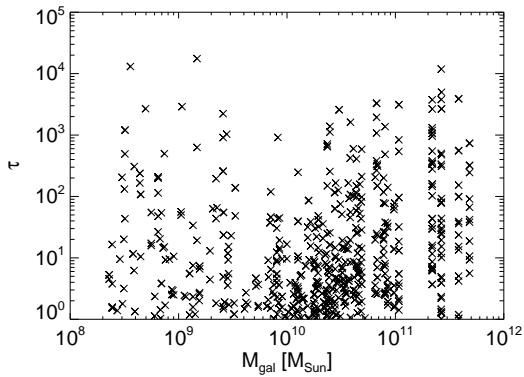}
\end{center}
\caption{Distribution of optical depths $\tilde{\tau}_{LL}$ of the \ll\ systems
  versus the mass of their associated galaxy.}
\label{fig:masstau}
\end{figure}

Figure~\ref{fig:masstau} shows the masses of the galaxies that the
\ll\ systems are associated with versus the systems'
optical depth. There is no strong dependence on galaxy mass for the
optical depth, meaning that the size of the galaxy does not appear to
strongly influence the opacity of the \ll\ system.

\begin{figure}[th]
\begin{center}
\epsscale{1.0}
\plotone{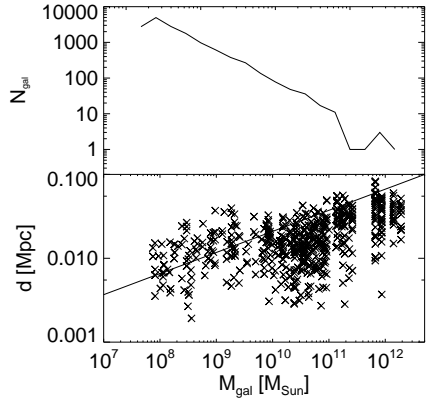}
\end{center}
\caption{\textit{Top Panel}: Number of galaxies per mass
  bin. \textit{Bottom Panel}: Distribution of distances of the \ll\
  systems from their associated galaxy with respect to the mass of
  that galaxy.}
\label{fig:massdist}
\end{figure}

The bottom panel of Figure~\ref{fig:massdist} is a graph of the
distance of the \ll\ system to its associated galaxy and the mass of
that galaxy. Overlaid on this correlation is the distance that
corresponds to the virial radius for each galaxy mass. The resolution
limit of our simulation is about $0.001\dim{Mpc}$, so that we resolve
all the distances to the galaxies. The top panel shows the number of
galaxies per mass bin for the same mass range as the bottom panel. It
is clear from this histogram that there are many more small galaxies
that have no \ll\ associated with them.

Combining the information from this graph, we find that most of the
\ll\ systems are within the virial radius of a galaxy, often a
relatively massive one. This means that they are either in orbit
around or infalling into a galaxy, and not isolated clumps of
neutral gas. Using the DENMAX classification of points in space
discussed in \S \ref{sec-simulations}, we find no \ll\ systems that
are unassociated or associated with unresolved galaxies.

\begin{figure}[th]
\begin{center}
\epsscale{1.0}
\plotone{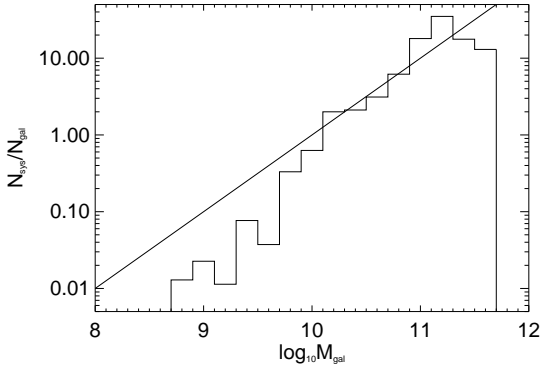}
\end{center}
\caption{Number of \ll\ systems per associated galaxy. Overlaid is the
  fit corresponding to an equal amount of \ll\ systems per unit log in
  galaxy mass.}
\label{fig:masshist}
\end{figure}

Figure~\ref{fig:masshist} extends the discussion from the previous
graph by showing the number of \ll\ systems associated with galaxies
of different masses. There are many more systems per galaxy in the
high galaxy mass range around $M_{gal}\sim10^{10}$ to
$M_{gal}\sim10^{11}$ than for lower masses. Since
$dN_{gal}/d\log(M)\sim M^{-1}$, the number of galaxies sharply
increases at low masses. In Figure~\ref{fig:masshist} the overlaid
power law ($\propto M^1$) shows the equal number of \ll\ systems per
logarithmic bin in mass of galaxies they are associated with. The
figure shows that only a small fraction of all \ll\ systems are
associated with galaxies of $M_{gal}<10^{10}M_\Sun$. However, this
effect might also be related to resolution effects, meaning that the
\ll\ systems are associated with galaxies of all mass ranges equally.

\section{Conclusions}

We show that the column density distribution of \ll\ systems at $z=4$ in the
cosmological simulations that agree with the SDSS measurements of the
\lya\ absorption at $z>5$ in the spectra of high redshift quasars has
the same shape as the observed distribution. The abundance of the \ll\
system is not necessarily in agreement with the data, but we manage to
achieve the agreement with both the observed column density
distribution of the \ll\ systems and with the mean transmitted flux in
the \lya\ forest by adjusting a single parameter - ionizing efficiency
- at $z<5$.

We find that, in the simulation that agrees with the observational
data, the lowest column density \ll\ systems are mainly illuminated by
the cosmological background, in agreement with previous findings of
\citet{me05} and \citet{s04}.

However, we also find that all \ll\ systems that are
resolved in our simulations ($N_{\HI}\la10^{20}\dim{cm}^2$) are highly
ionized and are {\it physically associated\/} with (are in orbit around or
falling into) the gravitational well of a large range of galaxy masses
($M_{tot}>10^{10}M_\Sun$). In other words, they appear as residual ``bits and
pieces'' of the galaxy formation process rather than self-shielded,
highly neutral dark matter ``minihalos'' ($M_{tot}<10^{9}M_\Sun$).

Because of the finite spatial and mass resolution of our simulations,
we can not exclude an existence of the second population of \ll\
systems composed of ``minihalos''. However, the fact that we are able
to fit both the \ll\ column density distribution and the mean
transmitted flux in the \lya\ forest at $z=4$ with one adjustable
parameter may indicate that the population of \ll\ systems arising in
``minihalos'' is not the dominant one. Other studies have also shown
that we can reproduce the observed \lya\ forest, so that the hidden
population cannot have any influence on the \lya\
forest. Cross-correlating \ll\ systems and galaxies could also help
determine whether this population is related to ``minihalos''.

\acknowledgements
We thank Jordi Miralda-Escud\'{e} and Lars Hernquist for their helpful
comments and suggestions.

This work was supported in part by the DOE and the NASA grant NAG
5-10842 at Fermilab, by the NSF grant AST-0134373,
and by the National Computational Science Alliance grant AST-020018N,
and utilized IBM P690 arrays at the National Center for Supercomputing
Applications (NCSA) and the San Diego Supercomputer Center (SDSC). KK
was supported by the Fellowship from the Department of Astrophysical
and Planetary Sciences of the University of Colorado.

\bibliography{LLpaperapr25}

\begin{thebibliography}{21}
\expandafter\ifx\csname natexlab\endcsname\relax\def\natexlab#1{#1}\fi

\bibitem[{{Bertschinger} \& {Gelb}(1991)}]{bg91}
{Bertschinger}, E. \& {Gelb}, J.~M. 1991, Computers in Physics, 5, 164

\bibitem[{{Blitz} {et~al.}(1999){Blitz}, {Spergel}, {Teuben}, {Hartmann}, \&
  {Burton}}]{blitz99}
{Blitz}, L., {Spergel}, D.~N., {Teuben}, P.~J., {Hartmann}, D., \& {Burton},
  W.~B. 1999, \apj, 514, 818

\bibitem[{{Fan} {et~al.}(2006){Fan}, {Strauss}, {Becker}, {White}, {Gunn},
  {Knapp}, {Richards}, {Schneider}, {Brinkmann}, \& {Fukugita}}]{fsbw06}
{Fan}, X., {Strauss}, M.~A., {Becker}, R.~H., {White}, R.~L., {Gunn}, J.~E.,
  {Knapp}, G.~R., {Richards}, G.~T., {Schneider}, D.~P., {Brinkmann}, J., \&
  {Fukugita}, M. 2006, \aj, in press

\bibitem[{{Gardner} {et~al.}(1997){Gardner}, {Katz}, {Hernquist}, \&
  {Weinberg}}]{gard97}
{Gardner}, J.~P., {Katz}, N., {Hernquist}, L., \& {Weinberg}, D.~H. 1997, \apj,
  484, 31

\bibitem[{{Gardner} {et~al.}(2001){Gardner}, {Katz}, {Hernquist}, \&
  {Weinberg}}]{gard01}
---. 2001, \apj, 559, 131

\bibitem[{{Gnedin}(2000)}]{g00a}
{Gnedin}, N.~Y. 2000, \apj, 535, 530

\bibitem[{{Gnedin}(2004)}]{g04}
---. 2004, \apj, 610, 9

\bibitem[{{Gnedin} \& {Fan}(2006)}]{gf06}
{Gnedin}, N.~Y. \& {Fan}, X. 2006, astroph/0603794

\bibitem[{{Katz} {et~al.}(1996){Katz}, {Weinberg}, {Hernquist}, \&
  {Miralda-Escude}}]{katz96}
{Katz}, N., {Weinberg}, D.~H., {Hernquist}, L., \& {Miralda-Escude}, J. 1996,
  \apjl, 457, L57+

\bibitem[{{Maloney} \& {Putman}(2003)}]{mal03}
{Maloney}, P.~R. \& {Putman}, M.~E. 2003, \apj, 589, 270

\bibitem[{{McDonald} {et~al.}(2002){McDonald}, {Miralda-Escud{\'e}}, \&
  {Cen}}]{macdon02}
{McDonald}, P., {Miralda-Escud{\'e}}, J., \& {Cen}, R. 2002, \apj, 580, 42

\bibitem[{{Miralda-Escud{\'e}}(2003)}]{me03}
{Miralda-Escud{\'e}}, J. 2003, \apj, 597, 66

\bibitem[{{Miralda-Escud{\'e}}(2005)}]{me05}
---. 2005, \apjl, 620, L91

\bibitem[{{Oort}(1966)}]{o66}
{Oort}, J.~H. 1966, Bull. Astr. Inst. Netherlands, 18, 421

\bibitem[{{Penton} {et~al.}(2002){Penton}, {Stocke}, \& {Shull}}]{pent02}
{Penton}, S.~V., {Stocke}, J.~T., \& {Shull}, J.~M. 2002, \apj, 565, 720

\bibitem[{{Petitjean} {et~al.}(1993){Petitjean}, {Webb}, {Rauch}, {Carswell},
  \& {Lanzetta}}]{pet93}
{Petitjean}, P., {Webb}, J.~K., {Rauch}, M., {Carswell}, R.~F., \& {Lanzetta},
  K. 1993, \mnras, 262, 499

\bibitem[{{Putman} {et~al.}(2003){Putman}, {Bland-Hawthorn}, {Veilleux},
  {Gibson}, {Freeman}, \& {Maloney}}]{put03}
{Putman}, M.~E., {Bland-Hawthorn}, J., {Veilleux}, S., {Gibson}, B.~K.,
  {Freeman}, K.~C., \& {Maloney}, P.~R. 2003, \apj, 597, 948

\bibitem[{{Schaye}(2006)}]{s04}
{Schaye}, J. 2006, astroph/0409137

\bibitem[{{Spergel} {et~al.}(2003){Spergel}, {Verde}, {Peiris}, \& {et
  al.}}]{svp03}
{Spergel}, D.~N., {Verde}, L., {Peiris}, H., \& {et al.} 2003, \apjs, 148, 175

\bibitem[{{Storrie-Lombardi} {et~al.}(1994){Storrie-Lombardi}, {McMahon},
  {Irwin}, \& {Hazard}}]{S-L94}
{Storrie-Lombardi}, L.~J., {McMahon}, R.~G., {Irwin}, M.~J., \& {Hazard}, C.
  1994, \apjl, 427, L13

\bibitem[{{Verschuur}(1969)}]{v69}
{Verschuur}, G.~L. 1969, \apj, 156, 771

\end{thebibliography}

\end{document}